\documentclass[letterpaper,12pt]{article}
\usepackage{dgjournal} 
\usepackage{newtxtext,newtxmath}
\usepackage{mathptmx}
\usepackage{graphics}
\usepackage[authoryear,comma,sectionbib]{natbib}
\usepackage{cancel}

\usepackage{geometry}
\usepackage{moreverb}
\usepackage{color}
\usepackage[titletoc,toc,title]{appendix}
\usepackage[normalem]{ulem}
\usepackage{soul}
\usepackage{amsmath}
\usepackage{amssymb}

\usepackage[colorlinks,bookmarksopen,bookmarksnumbered,citecolor=red,urlcolor=red]{hyperref}

\newcommand\BibTeX{{\rmfamily B\kern-.05em \textsc{i\kern-.025em b}\kern-.08em
T\kern-.1667em\lower.7ex\hbox{E}\kern-.125emX}}

\newcommand{\p}{\mathbb{P}}
\newcommand{\E}{\mathbb{E}}

\newcommand{\1}{\mathbb{I}}

\newcommand{\bC}{{\boldsymbol C}}
\newcommand{\bV}{{\boldsymbol V}}
\newcommand{\bcV}{{\boldsymbol {\cal V}}}
\newcommand{\bv}{{\boldsymbol v}}
\newcommand{\bW}{{\boldsymbol W}}
\newcommand{\bw}{{\boldsymbol w}}
\newcommand{\cM}{{\mathcal M}}
\newcommand{\eps}{{\varepsilon}}

\makeatletter
\newcommand*{\indep}{%
  \mathbin{%
    \mathpalette{\@indep}{}%
  }%
}
\newcommand*{\nindep}{
  \mathbin{
    \mathpalette{\@indep}{\not}    
  }
}

\newcommand*{\@indep}[2]{%
  \sbox0{$#1\perp\m@th$}
  \sbox2{$#1=$}
  \sbox4{$#1\vcenter{}$}
  \rlap{\copy0}
  \dimen@=\dimexpr\ht2-\ht4-.2pt\relax
  \kern\dimen@
  {#2}%
  \kern\dimen@
  \copy0 
} 
\makeatother

\usepackage{tikz}
\usetikzlibrary{arrows,shapes}
\tikzstyle{var}=[draw,circle,thick, text width=4mm, inner sep = 3pt]
\tikzstyle{varr}=[draw,circle,thick]
\tikzstyle{varf}=[draw=none,fill=none]
\tikzstyle{varCond}=[draw,rectangle,rounded corners=3pt, fill=gray, minimum width=2em,minimum height=2em]
\tikzstyle{edge} = [draw,thick,->,>=latex]
\tikzstyle{edge2} = [draw,thick,-]
\tikzstyle{edge3} = [draw,dashed,->,>=latex] 

\tikzset{
semi/.style={
  semicircle,
  draw,
  minimum size=2em
  }
}

\tikzstyle{varcounterfactual}=[draw, ellipse] 
\tikzstyle{varright} = [draw, semi, shape border rotate=270] 
\tikzstyle{varleft} = [draw,  semi,shape border rotate=90]

\vfuzz \hfuzz
\begin{document}
\begin{center}
{\LARGE Causal inference to detect selection bias}  \\
\vspace{0.55cm}
{\LARGE in road safety epidemiology} \\ 
\end{center}

M. Dufournet$^{\left(1\right)}$, E. Lanoy$^{\left(2\right),\left(3\right)}$, J.L. Martin$^{\left(1\right)}$, V. Viallon$^{\left(1\right)}$ \\

\vspace{2cm}
\noindent{\small $^{\left(1\right)}$ Univ  Lyon, Universit\'e Claude Bernard Lyon 1, Ifsttar, UMRESTTE, UMR T\_9405, F-69373 LYON \\
$^{\left(2\right)}$ Gustave Roussy, Universit\'e Paris-Saclay, Service de biostatistique et d'\'epid\'emiologie, Villejuif, F-94805 \\
$^{\left(3\right)}$ Universit\'e Paris-Saclay, Univ. Paris-Sud, UVSQ, CESP, INSERM, Villejuif, F-94085} \\

\begin{abstract}
In the field of road safety, it is common to use responsibility analyses to assess the effect of a given factor on the risk of being responsible for an accident, among drivers involved in an accident only. 
Even if this design is now widely adopted in the field, the question of selection bias is often raised. The structural Causal Model framework now provides  valuable tools to assess causal effects from observational data and identify selection bias. In this article, we briefly review recent results regarding the recoverability of causal effects from selection biased data, and apply them to the case of responsibility analyses. Our objective is to formally determine whether causal effects can be unbiasedly estimated through this type of analyses, when available data are restricted to severe accidents, as it is commonly the case in practice. However, because speed has a direct effect on the severity of the accident, we show that causal odds-ratios are not estimable from responsibility analyses. We present numerical results to illustrate our argument, the magnitude of the bias and to discuss recent results from real data. \\ 


\end{abstract}

\newpage
\section{Introduction} 

Research into road safety must recognise causality. Preventing road crashes, in the interest of public health and safety, implies identifying the underlying causes of crashes. In the field, main causes of the occurrence and severity of road crashes due to human behavior have been established: driving under the influence of alcohol or drugs, inappropriate speed, distracted driving, etc. Questions asked by decision-makers now mostly concern the magnitude of their causal effects, as well as the burden of deaths or victims attributable to these various causes of accident \citep{Cummings_2006}. \\
In order to measure causal effects, the ideal designs are interventional studies such as the randomized clinical trial \citep{mill_system_1843}. However, interventional studies are not always possible in epidemiology or clinical research, either for ethical, costs or other technical reasons  \citep{baiocchi_instrumental_2014}.  Consequently, it is common to only have access to observational data. In this case, two main sources of bias often arise: confounding and selection bias \citep{hernan2002causal,greenland2003quantifying,elwert2013graphical,elwert2014endogenous}.\\
In the case of road safety epidemiology, no interventional study can be performed for obvious ethical reasons. In addition, data are usually only available when the outcome of interest, the accident, occurred. Indeed, data are restricted to drivers and vehicles involved in road accidents only, and often to severe road accidents only (e.g., injury or fatal accidents). This extreme selection of data precludes the estimation of the effect of any exposure on the risk of car accident. Therefore, the first step before estimating causal effects is to define appropriate cases and controls. Two closely related approaches have been adopted to deal with this issue: quasi-induced exposure \citep{Stamatiadis_1997} and responsibility analysis \citep{Smith_1951, Perchonok_1978, Terhune_1986, brubacher2014culpability}. The general idea is to assess the causal effect of a given factor on the risk of being responsible for an accident, or a severe accident, among involved drivers. The two approaches first rely on the assessment of the responsibility of each driver involved in crashes, usually from police reports. They then consist in comparing responsible drivers with non-responsible drivers among involved drivers. The underlying assumption is that non-responsible drivers represent a random sample of the general driving population that was "selected" to crash by circumstances beyond their control and therefore have the same risk factor profile as other drivers on the road at the same time \citep{brubacher2014culpability, Wahlberg_2007}. A standard claim is that "if this randomness assumption is met, then the risk estimate derived from a responsibility analyses would be expected to be similar to that from a standard case-control study" \citep{brubacher2014culpability, Wahlberg_2007}. Thus, responsibility analyses would allow researchers to unbiasedly estimate the causal effect of any exposure on the risk of being responsible for an accident, by restricting the analysis to drivers involved in an accident only.  Even if induced exposure and responsibility analyses are now widely adopted in the field \citep{Asbridge_2013, Salmi_2014, Wahlberg_2009}, some authors have raised questions about the validity of these approaches and the potential presence of selection bias. Indeed, the randomness assumption is questionable since the non-responsible drivers can be very different from the general driving population \citep{Sanghavi_2013}. Consequently, responsibility analyses could lead to biased estimations, even if relevant confounders have been taken into account. Nethertheless, the question is still open and responsibility analyses are rarely discussed and challenged. \\
Our objective is to answer this question by determining whether the causal effects of a given factor can be unbiasedly estimated through responsibility analyses. The singularity of our work is to use the causal reasoning, the Structural Causal Model (SEM) framework, and recent results regarding the recoverability of causal effects in the presence of selection mechanism. Our results enable us to close the debate, unfortunately in favor of biased estimations, but they also give interesting leads to interpret and discuss estimations derived from responsibility analyses. \\
The article is organized as follows. In Section \ref{sec:SelBias} below, we briefly recall some basics of the SCM framework and recent results regarding \textit{the recoverability} of causal effects in the presence of selection bias. In Section \ref{sec:appli},  the application to  responsibility analyses is presented. In Section \ref{sec:NumIllustr}, we present results from a simple numerical analysis to illustrate the possible magnitude of bias, and we discuss recent results from real data. In Section \ref{sec:ccl}, we conclude with leads for future research.

\section{Causal inference in the presence of selection mechanism} \label{sec:SelBias}

In this section, we present the Structural Causal Model (SCM) framework and study the recoverability of causal effects in the general context of selection biased data. \\

\noindent We consider the scenario where $Y$ denotes a dichotomous outcome, $X$ a binary exposure of interest, and $\bW$ is a vector of additional categorical variables. We will denote by $\bcV$ the set of observable variables $\left(X,\bW,Y\right)$. The causal model $\cM$ leading to $Y$ can be graphically represented by a DAG $G$ (Directed Acyclic Graph) \citep{Pearl_1995,greenland_causal_1999,glymour_causal_2008}. In the SCM framework, the DAG is associated to a set of structural functions, each corresponding to one of the covariates in the DAG. See Appendix \ref{ap:A0} for more details. This set of equations allows the definition of $Y_x$, the counterfactual outcome that would have been observed in the counterfactual world where exposure would have been set to $X=x$, for $x \in \lbrace 0,1 \rbrace$. Then, causal effects can be precisely defined. In particular, in the simple case considered here where both $X$ and $Y$  are binary, the average causal total effect (ACE) of $X$ on $Y$ is defined as:
\begin{equation}
{\rm ACE}  = \E\left(Y_{1} - Y_{0}\right)  = \p\left(Y_1 = 1\right) - \p\left(Y_0 = 1\right). \label{eq:ACE1} \\ 
\end{equation}
Here, the causal effect is defined on the excess risk scale, but causal risk ratio or odds-ratio can be defined similarly.\\

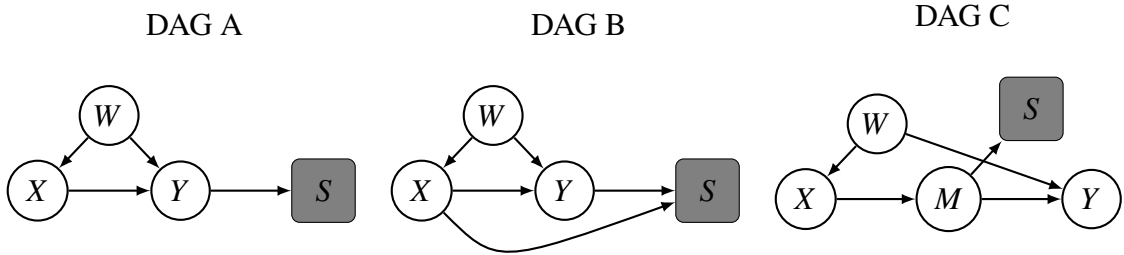
\begin{figure}[!h]
\begin{minipage}{5cm}
\begin{center} DAG A \end{center}
\begin{tikzpicture}[scale=1, auto,swap]
\node[varr] (X)at(0,0){$X$};
\node[varr] (Y)at(1.9,0){$Y$};
\node[varCond] (S)at(3.8,0){$S$};
\node[var] (W) at (0.95,1) {$\bW$};
\draw[edge] (W)--(X);
\draw[edge] (W)--(Y);
\draw[edge] (X)--(Y);
\draw[edge] (Y)--(S);
\phantom{\draw[edge] (X).. controls (1,-1) ..(S);}
\end{tikzpicture}
\end{minipage}
\begin{minipage}{5cm}
\begin{center} DAG B \end{center}
\begin{tikzpicture}[scale=1, auto,swap]
\node[varr] (X)at(0,0){$X$};
\node[varr] (Y)at(1.9,0){$Y$};
\node[varCond] (S)at(3.8,0){$S$};
\node[var] (W) at (0.95,1) {$\bW$};
\draw[edge] (W)--(X);
\draw[edge] (W)--(Y);
\draw[edge] (X)--(Y);
\draw[edge] (Y)--(S);
\draw[edge] (X).. controls (1,-1) ..(S);
\end{tikzpicture}
\end{minipage}
\begin{minipage}{5cm}
\begin{center} DAG C \end{center}
\begin{tikzpicture}[scale=1, auto,swap]
\node[varr] (X)at(0,0){$X$};
\node[varCond] (S)at(3,1.2){$S$};
\node[varr] (M)at(1.9,0){$M$};
\node[varr] (Y)at(3.8,0){$Y$};
\node[var] (W) at (0.95,1) {$\bW$};
\draw[edge] (W)--(X);
\draw[edge] (W)--(Y);
\draw[edge] (X)--(M);
\draw[edge] (M)--(Y);
\draw[edge] (M)--(S);
\phantom{\draw[edge] (X).. controls (1,-1) ..(Y);}
\end{tikzpicture}
\end{minipage}
\caption{Examples of DAGs in the presence of selection bias}
\label{fig1}
\end{figure}

Causal inference is mainly concerned with the identification of $\p\left(Y_x = y\right)$ for $\left(x,y\right) \in \lbrace 0,1 \rbrace$ and then of the causal effects, from the distribution of the observed variables $\p\left(\bcV=.\right)$ \citep{Bareinboim_Tian_2015}. When there is no selection bias, criteria such as the back-door and the front-door criteria ensure the identifiability of causal effects \citep{Pearl_1995,Pearl_2009}. When selection bias is present, the binary variable $S$ indicating inclusion in the study has to be added to $G$, leading to a new DAG $G_s$. It is standard to represent $S$ in a specific way in $G_s$ to emphasize that $S$ does not play any role in the causal model described by $G$ but plays a role in the selection process since data are only available for individuals for whom $S=1$ (see Figure \ref{fig1}). Then, the question is the recoverability of $\p\left(Y_x = y\right)$ for $\left(x,y\right) \in \lbrace 0,1 \rbrace$ in terms of the observable distribution $\p\left({\boldsymbol {\cal V}}=.| S=1\right)$ \citep{Bareinboim_Pearl_2012, Bareinboim_Tian_2015}. Recoverability can be seen as a generalization of identifiability in the presence of selection bias. Whether this selection may result in biased estimation of the causal effects of interest depends on the structure of $G_s$, and in particular on the arrows pointing to or emanating from $S$ in $G_s$ \citep{Hernan_2004}. Two types of results allow us to determine whether the causal effect is recoverable depending on the structure of the DAG $G_s$. First, \cite{Bareinboim_Tian_2015} define necessary and sufficient conditions ensuring the recoverability of the quantity $\p\left(Y_x = y\right)$, and as a byproduct the recoverability of the causal effects (causal excess risk, causal relative risk, causal odds-ratio), in the presence of selection. Second, under additional assumptions, causal odds-ratios are recoverable from selection biased data in specific situations where the distribution $\p\left(Y_x = y\right)$ and other causal effects are not \citep{Bareinboim_Pearl_2012}. For illustration, and motivated by the context of responsibility analyses, we mostly focus on situations where inclusion in the study depends on the outcome $Y$, as in DAGs A and B of Figure \ref{fig1}. We also consider the case where $S$ is a descendant of an intermediate variable (\rm{i.e.}, a descendant of $X$), as in DAG C of Figure \ref{fig1}.

\subsection{Recoverability of $\p\left(Y_x = y\right)$ in the presence of selection mechanism}

A central result for the recoverability of $\p\left(Y_x = y\right)$ in the presence of selection is stated by \cite{Bareinboim_Tian_2015}. As above, denote by $G$ the DAG of interest, composed of the observable variables $\bcV$, and by $G_s$ the DAG obtained after the addition of the selection variable $S$ in $G$.  For any $\bC\subseteq \bcV$, further define $G_{\bC}$ the subgraph of DAG $G$ composed of the variables in $\bC$ only. For any $V_i\in \bcV$, further denote by $An\left(V_i\right)_G$ the union of $V_i$ and the ancestors of $V_i$ in DAG $G$. Then, Theorem 2 in \cite{Bareinboim_Tian_2015} states that $\p\left(Y_x = y\right)$ is recoverable from selection biased data if and only if 
$$ {\rm \left(C.1\right)}\quad\quad\quad\quad An\left(Y\right)_{G_{\bcV \setminus X}}  \cap An\left(S\right)_{G_s} = \emptyset.$$ 

Condition (C.1) above is violated in the three examples depicted in DAGs A, B and C of Figure \ref{fig1}.  For instance, we have $Y\in An\left(Y\right)_{G_{\bcV \setminus X}} \cap An\left(S\right)_{G_s}$ in DAGs A and B, and $M\in An\left(Y\right)_{G_{\bcV \setminus X}} \cap An\left(S\right)_{G_s}$ in DAG C. Therefore, $\p\left(Y_x = y\right)$ is not recoverable under these types of selection. Examples of DAGs ensuring that $\p\left(Y_x = y\right)$ is recoverable from selection biased data can be found in \cite{Bareinboim_Tian_2015}.  

\subsection{Recoverability of causal odds-ratios in the presence of selection bias}\label{Sec:ORSelBias}

Assume that $\bW$ is a vector of confounders between $X$ and $Y$, as in DAGs A, B and C. Here, we consider the recoverability of $\bw$-specific causal odds-ratio
$$COR\left(X, Y | \bW=\bw\right) = \frac{\p\left(Y_1=1 |\bW=\bw\right)/\p\left(Y_1=0 |\bW=\bw\right)}{\p\left(Y_0=1 |\bW=\bw\right)/\p\left(Y_0=0 |\bW=\bw\right)}.$$
The $\bw$-specific causal odds-ratio is the causal odds-ratio in the stratum of the population defined by $\bW=\bw$. When $\bW$ contains a set of sufficient confounders, conditional ignorability $Y_x \indep X | \bW$ holds \citep{Pearl_2000,vanderweele_minimal_2009,vanderweele_new_2011}, so that $\p\left(Y_x=1 | \bW=\bw\right)=\p\left(Y=1 | X=x, \bW=\bw\right)$. Then, $\bw$-specific causal odds-ratio equals the adjusted odds-ratio $$OR\left(X, Y | \bW=\bw\right) = \frac{\p\left(Y=1 | X=1, \bW=\bw\right)/\p\left(Y=0 | X=1, \bW=\bw\right)}{\p\left(Y=1 | X=0, \bW=\bw\right)/\p\left(Y=0 | X=0, \bW=\bw\right)}.$$

Following Definition 2 of \cite{Bareinboim_Pearl_2012}, $OR\left(X, Y | \bW=\bw\right)$, and then in our case $COR\left(X, Y | \bW=\bw\right)$,  are recoverable from selection biased data if the assumptions embedded in the DAG renders it expressible in terms of the observable distribution $\p\left(\bcV=.| S=1\right)$. 
The symmetry of the odds-ratio $OR\left(X, Y | \bW=\bw\right) = OR\left(Y, X | \bW=\bw\right)$ makes it recoverable in specific situations where $\p\left(Y_x=y\right)$ is not. More precisely, Theorem 1 in \cite{Bareinboim_Pearl_2012}, or Corollary 4 in \cite{didelez_graphical_2010}, states that $OR\left(X, Y | \bW=\bw\right)$  is recoverable from selection biased data if and only if 
$$ {\rm \left(C.2\right)}\quad\quad\quad\quad X\indep S |  \left(Y, \bW\right) \text{ or } Y\indep S |  \left(X, \bW\right). $$ 
This condition holds under DAG A, but is not guaranteed under DAGs B and C. Consequently, $OR\left(X, Y | \bW=\bw\right)$ and then $COR\left(X,Y|\bW=\bw\right)$ are recoverable under DAG A, while they are not under DAGs B and C. \\

\noindent In this paragraph, we focus on DAGs A and B, \rm{i.e.} on situations where inclusion depends on the outcome $Y$, as in responsibility analyses. 
Through these two DAGs, we can note that the $\bw$-specific causal odds-ratio is not recoverable as soon as the selection is affected by the outcome and the exposure. This situation is related to the general phenomenon called collider bias  \citep{greenland2003quantifying,rothman_modern_2008}. We will see that it is at play in responsibility analyses.

\section{Application to responsibility analyses} \label{sec:appli} 

We can now check whether the conditions to recover causal total effects in the presence of selection bias hold in the particular setting of responsibility analyses. Under-reporting of crashes is a well-recognised
problem; the more serious the crash, the more likely it is to be recorded \citep{amoros_estimation_2008}. As a result available data are often restricted to, or at least mostly concern, severe accidents. Here, we will consider the estimation of the causal total effect of a given exposure on the ``responsibility'' of a severe accident among drivers involved in a severe car crash. \\

\subsection{Formalization of responsibility analyses}\label{sec:Resp} 

\noindent First, let us formalize what ``responsibility'' means in responsibility analyses. 
In this context, the ``responsibility'' of a driver is not entirely driven by the determination of a legal fault. A driver involved in a crash is considered as responsible of this crash if he committed a driving fault (lane departure, failure to obey traffic signs, driving against traffic, etc), which should trigger the crash. Note that alcohol consumption or cannabis intoxication are traffic violation (illegal and frowned upon) but they are not driving faults. Rather, they are causes of driving faults. The responsibility, which is assigned after the crash from police reports, is a measure of the driving fault, which takes place before the crash. We will denote by $F$ the binary variable indicating whether the driver commits a driving fault. Observe that this variable is defined for all drivers, not only those involved in a crash. Further denote by $A$ the binary variable indicating whether this driver is involved in a severe accident. We can therefore consider the binary variable $R$ indicating whether the driver is responsible for a severe accident. So, $R = F\times A$, and we have:

\begin{displaymath}
\left\{ \begin{array}{ll}
R=1 & \textrm{if and only if } A=1 \textrm{ and } F=1 \\
R=0 & \textrm{if } A=1 \textrm{ and } F=0 \\
R=0 & \textrm{if } A=0 \textrm{ even if } F=1. \\
\end{array} \right. \\
\end{displaymath}

\vspace{0.25cm}
\noindent First of all, we will focus on the recoverability of causal effects on the responsibility of a severe accident $R$. But, we will see in a second step that it can be sometimes useful to consider the recoverability of the $\bw$-specific causal odds ratio on the driving fault $F$ which can serve as a basis to approximate the $\bw$-specific causal odds ratio on the responsibility of a severe accident $R$. \\

\noindent The DAG $G$ below represents a simplified causal mechanism which leads to a severe accident from one driver's point of view. As already mentioned, an accident (severe or not) generally occurs because of a driving fault. That is the reason why there is an arrow from $F$ to $A$. This driving fault is potentially caused by the considered exposure $X$ (alcohol consumption, cannabis intoxication, cell-phone use...), hence the arrow from $X$ to $F$. $X$, $F$ and $A$ may be affected by many confounders, like age, gender, daytime, period of the year, speed limit etc. We denote by $\bW$ the set of all confounders. Most often, the exposure $X$ is related to speed. We denote by $V$ the binary variable indicating whether a given driver drives at high or inappropriate speed. We first consider a case where $X$ has an impact on high speed $V$, such as alcohol, so there is an arrow from $X$ to $V$. $V$ has an impact both on driving fault $F$ and on severe accident $A$. Indeed, on the one hand, high speed has an impact on driving performances since high speed increases the risk of being unable to fit his driving to circumstances. Consequently, it increases the risk to commit a driving fault $F$. Driving at an inappropriate speed can also be considered as a driving fault. On the other hand, high speed has an impact on accident severity: the higher the speed, the more serious the crash. As other variables, $V$ is also affected by $\bW$. Finally, $R$ depends on $F$ and $A$, because $R=F \times A$. \\

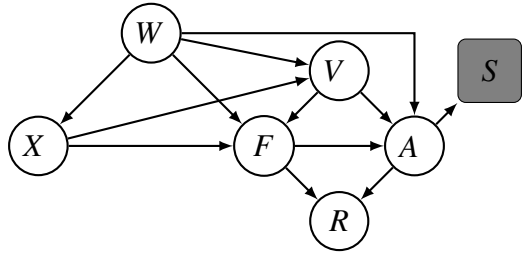
\begin{figure}[!h]
\vspace{0.3cm}
\begin{center}
\begin{tikzpicture}[scale=1, auto,swap]
\node[var] (X)at(0,0){$X$};
\node[varr] (F)at(3,0){$F$};
\node[var] (A)at(5,0){$A$};
\node[var] (W) at (1.5,1.5) {$\bW$};
\node[var] (V) at (4,1) {$V$};
\node[varr] (R) at (4,-1) {$R$};
\node[varCond] (S)at(6,1){$S$};
\draw[edge] (W)--(X);
\draw[edge] (W)--(V);
\draw[edge] (W)--(F);
\draw[edge] (X)--(F);
\draw[edge] (X)--(V);
\draw[thick,->,>=latex] (W) -| (A);
\draw[edge] (F)--(A);
\draw[edge] (F)--(R);
\draw[edge] (A)--(R);
\draw[edge] (V)--(F);
\draw[edge] (V)--(A);
\draw[edge] (A)--(S);
\end{tikzpicture}
\end{center}
\caption{DAG $G$ representing the causal mechanism leading to the occurence of a severe accident $A$.}
\label{fig20}
\end{figure}

\noindent In Section \ref{sec:Resp(ii)}, we apply the principles on the DAG illustrated by Figure \ref{fig20}, \rm{i.e} when data are available for severe crashes only ($A=1$) and the considered exposure $X$ causes high speed $V$. In this case, selection is affected by both the outcome and the exposure, so we will see that the causal effect is not recoverable. In Section \ref{sec:Resp(i)} we will describe other situations where the causal effect could be well approximated. \\

\subsection{Recoverability of causal effects in responsibility analyses} \label{sec:Resp(ii)}

In the DAG of Figure \ref{fig20}, $A$ represents the occurence of a severe crash, on which selection depends in the sense that $S=1 \Rightarrow A=1$. \\

\noindent Because  $A\in An\left(R\right)_{G_{\bV \setminus X}} \cap An\left(S\right)_{G_s}$, results presented in Section \ref{sec:SelBias} indicate that $\p\left(R_x=r\right)$ is not recoverable in this context, for $\left(x,r\right)\in\{0,1\}$.  Regarding $\bw$-specific odds-ratios, $COR\left(X, R| \bW=\bw\right)$  is not recoverable since neither $X\indep S| \left(R, \bW\right)$ nor $R\indep S| \left(X, \bW\right)$ is guaranteed. On the one hand, the set $\left(R, \bW\right)$ does not $d$-separate all the paths from $X$ to $S$ since it does not $d$-separate the path $X \longleftarrow V \longrightarrow A \longrightarrow S$. On the other hand, the set $\left(X, \bW\right)$ does not $d$-separate all the paths from $R$ to $S$ since $\left(X, \bW\right)$ does not $d$-separate the path $R \longleftarrow F \longrightarrow A \longrightarrow S$. As already mentioned, it is sometimes useful to consider the recoverability of $COR\left(X,F| \bW=\bw\right)$, but $COR\left(X,F| \bW=\bw\right)$ is not recoverable here since neither $X\indep A| \left(F, \bW\right)$ nor $F\indep A| \left(X, \bW\right)$ generally holds.\\
\noindent Note that a more realistic assumption may be that the selection depends not only on the accident, but on $\bW$ too. We can imagine that the selection depends on the type of road, since we observe more accidents on highway for a given level of severity. But even if the selection depends on $A$ and $\bW$, the conclusions remain identical to the ones exposed here. \\ 

\noindent To recap, in the situation considered here where there is an arrow between $X$ and $V$, and another one between $V$ and $A$, the causal effect of interest is not recoverable. For instance, the estimation of the causal effect of alcohol in responsibility analyses focused on fatal accident is biased. 

In other words, the estimable adjusted odds-ratio $OR\left(X,R| \bW=\bw, A=1\right)$, which equals 
\begin{align*}
 \frac{\p\left(R=1 | X=1, \bW=\bw, A=1\right)/\p\left(R=0 | X=1, \bW=\bw, A=1\right)}{\p\left(R=1 | X=0, \bW=\bw, A=1\right)/\p\left(R=0 | X=0, \bW=\bw, A=1\right)}.
\end{align*}
\noindent is not equal to $COR\left(X,R|\bW=\bw\right)$. It is important to further note that $OR\left(X,R|\bW=\bw, A=1\right)$ is not causal even inside the subpopulation $\{A=1\}$ \citep{frangakis2002principal}, that is, $OR\left(X,R|\bW=\bw, A=1\right) \neq COR\left(X,R|\bW=\bw, A=1\right)$, with
\begin{align*}
COR\left(X,R| \bW=\bw, A=1\right) & = \frac{\p\left(R_1=1 | \bW=\bw, A=1\right)/\p\left(R_1=0 | \bW=\bw, A=1\right)}{\p\left(R_0=1 | \bW=\bw, A=1\right)/\p\left(R_0=0 | \bW=\bw, A=1\right)}. 
\end{align*}
See Appendix \ref{ap:A03} for more details.  

\noindent Consequently, adjusted odds-ratios $OR\left(X, R | \bW=\bw, A=1\right)$ available in responsibility analyses have to be interpreted with caution when available data are restricted to severe accidents. They correspond neither to causal quantities generalizable to people outside the population of injured drivers, nor to causal quantities inside this population.\\

\subsection{Recoverability of causal effect in other situations}\label{sec:Resp(i)}

In this part, we study whether causal odds-ratios $COR\left(X,R|\bW=\bw\right)$ could be approximated if some paths were absent in the DAG of Figure \ref{fig20}. \newline
We consider three situations (See Figure \ref{fig7}): the case $\left(i\right)$ where $X$ would not be a cause of $A$, the case $\left(ii\right)$ where $X$ would not be a cause of $V$ and $V$ would not be a cause of $F$ and the case $\left(iii\right)$ where $X$ would not be a cause of $V$. 

\begin{figure}[!h]
\begin{minipage}{7cm}
\begin{center} \textit{(i)}  $V$ would not be a direct cause of $A$  \end{center}
\begin{center}
\begin{tikzpicture}[scale=1, auto,swap]
\node[varr] (X)at(0,0){$X$};
\node[varr] (F)at(3,0){$F$};
\node[varr] (A)at(5,0){$A$};
\node[varCond] (S)at(6,1){$S$};
\node[var] (W) at (1.5,1.5) {$\bW$};
\node[varr] (R) at (4,-1) {$R$};
\node[var] (V) at (4,1) {$V$};
\draw[edge] (W)--(X);
\draw[edge] (W)--(F);
\draw[thick,->,>=latex] (W) -| (A);
\draw[edge] (W)--(V);
\draw[edge] (X)--(F);
\draw[edge] (F)--(A);
\draw[edge] (F)--(R);
\draw[edge] (A)--(R);
\draw[edge] (A)--(S);
\draw[edge] (V)--(F);
\draw[edge] (X)--(V);
\end{tikzpicture}
\end{center}
\end{minipage}
\hspace{1cm}
\begin{minipage}{7cm}
\begin{center} \textit{(ii)} $X$ would not be a cause of $V$ and $V$ would not be a cause of $F$ \end{center}
\begin{center}
\begin{tikzpicture}[scale=1, auto,swap]
\node[varr] (X)at(0,0){$X$};
\node[varr] (F)at(3,0){$F$};
\node[varr] (A)at(5,0){$A$};
\node[varCond] (S)at(6,1){$S$};
\node[var] (W) at (1.5,1.5) {$\bW$};
\node[var] (V) at (4,1) {$V$};
\node[varr] (R) at (4,-1) {$R$};
\draw[edge] (W)--(X);
\draw[edge] (W)--(F);
\draw[thick,->,>=latex] (W) -| (A);
\draw[edge] (W)--(V);
\draw[edge] (X)--(F);
\draw[edge] (F)--(A);
\draw[edge] (F)--(R);
\draw[edge] (A)--(R);
\draw[edge] (A)--(S);
\draw[edge] (V)--(A);
\end{tikzpicture}
\end{center}
\end{minipage}
\begin{minipage}{17cm}
\begin{center} \textit{(iii)} $X$ would not be a cause of $V$ \end{center}
\begin{center}
\begin{tikzpicture}[scale=1, auto,swap]
\node[varr] (X)at(0,0){$X$};
\node[varr] (F)at(3,0){$F$};
\node[varr] (A)at(5,0){$A$};
\node[varCond] (S)at(6,1){$S$};
\node[var] (W) at (1.5,1.5) {$\bW$};
\node[var] (V) at (4,1) {$V$};
\node[varr] (R) at (4,-1) {$R$};
\draw[edge] (W)--(X);
\draw[edge] (W)--(F);
\draw[thick,->,>=latex] (W) -| (A);
\draw[edge] (W)--(V);
\draw[edge] (X)--(F);
\draw[edge] (F)--(A);
\draw[edge] (F)--(R);
\draw[edge] (A)--(R);
\draw[edge] (A)--(S);
\draw[edge] (V)--(F);
\draw[edge] (V)--(A);
\end{tikzpicture}
\end{center}
\end{minipage}
\caption{Other possible DAGs in responsibility analyses} \label{fig7}
\end{figure}

\newpage
\noindent In cases $\left(i\right)$ and $\left(ii\right)$, the conditional independence $X \indep A |\left(F,W\right)$ holds, so that $COR\left(X,R|\bW=\bw\right)$ can be well approximated by $COR\left(X,F|\bW=\bw\right)$. 
\noindent Indeed, although neither $\p\left(R_x=r\right)$ nor $COR\left(X,R| \bW=\bw\right)$ is recoverable (as in Section \ref{sec:Resp(i)}), $COR\left(X, F| \bW=\bw\right)$ is recoverable because $X\indep S| \left(F, \bW\right)$, which is implied by $X\indep A| \left(F, \bW\right)$. In other words, $COR\left(X, F| \bW=\bw\right) = OR\left(X,F|\bW=\bw, A=1\right) =  OR\left(X,R|\bW=\bw, A=1\right)$, which equals
\begin{eqnarray*}
&&\frac{\p\left(F=1 | X=1, \bW=\bw, A=1\right)/\p\left(F=0 | X=1, \bW=\bw, A=1\right)}{\p\left(F=1 | X=0, \bW=\bw, A=1\right)/\p\left(F=0 | X=0, \bW=\bw, A=1\right)} \\
 &=& \frac{\p\left(R=1 | X=1, \bW=\bw, A=1\right)/\p\left(R=0 | X=1, \bW=\bw, A=1\right)}{\p\left(R=1 | X=0, \bW=\bw, A=1\right)/\p\left(R=0 | X=0, \bW=\bw, A=1\right)},
\end{eqnarray*}
where the last equality comes from the fact that $F= R$ in the subpopulation $\{A=1\}$.  
In addition, if $\p\left(F_x=1|  \bW=\bw\right)$ and $\p\left(R_x=1|  \bW=\bw\right)$ are both small for $x\in\{0,1\}$, it can be shown that 
\begin{equation}
COR\left(X, R| \bW=\bw\right) \approx COR\left(X, F| \bW=\bw\right).  \label{eq:CORCRR}
\end{equation} 
\noindent See Appendix \ref{ap:CORCRR} for the proof of this result. Therefore, if $\p\left(F_x=1|  \bW=\bw\right)$ and $\p\left(R_x=1|  \bW=\bw\right)$ are small, $COR\left(X, R| \bW=\bw\right)$ is approximately recoverable in case $\left(i\right)$. \\

\noindent Unfortunately, cases $\left(i\right)$ and $\left(ii\right)$ are not realistic. In case $\left(i\right)$, the assumption that high speed $V$ would not be a cause of a severe accident does not seem reasonable. The absence of an arrow from $V$ to $A$ might be more plausible if $A$ represents the occurence of an accident, irrespective to its severity. In this situation, and considering driving at high or inappropriate speed as a driving fault $F$, then it might be argued that there is no arrow from $V$ to $A$. The absence of this arrow would still remain questionable. In case $\left(ii\right)$, the absence of an arrow from $V$ to $F$ does not seem plausible. Indeed, high speed increases the risk of loss of control and consequently the risk to commit a driving fault $F$.

\noindent Then case $\left(iii\right)$ appears as the most plausible. However, condition $X \indep A | \left(F,W\right)$ does not hold in this case since $\left(F,\bW\right)$ does not block the path $X \longrightarrow F \longleftarrow V \longrightarrow A$. We have $X \indep A | \left(F,W, V\right)$ so that $COR\left(X,R|\bW=\bw, V=v\right)$ is recoverable under case $\left(iii\right)$. It requires $V$ to be observed, which is rarely the case in practice.

\section{Numerical illustration}\label{sec:NumIllustr}

Above, we have shown that the absence of bias in responsibility analyses is only guaranteed if $X \indep A |\left(F,W\right)$ holds, which is not the case under realistic DAGs. However, for practitioners, it is useful to quantify the magnitude of an identified bias. Here, we present results from a simple numerical study to give a first quantification of the magnitude of bias induced by responsibility analyses in a simple causal model, and where $X$ could represent alcohol consumption. Note that we do not consider any sampling properties and hence do not simulate data. Rather we compare theoretical quantities under a given choice of a joint distribution of $\left(X,V,F,A,R,\bW\right)$ \citep{didelez2010assumptions}. 

\subsection{Full model} \label{sec:DataGen}

We consider a model consistent with the DAG of Figure \ref{fig20}. More precisely, our causal model is obtained by specifying the structural functions $f_F$, $f_A$, $f_X$, and $f_V$ as well as the distributions of the disturbances $\eps_F$, $\eps_A$, $\eps_X$, and $\eps_V$. 

\noindent Denote the indicator function by $\1[\cdot]$. Define four independent random variables $\eps_F$, $\eps_A$, $\eps_X$, and $\eps_V$  distributed according to a uniform distribution over the interval $[0,1]$. For any given  $p_X \in\left(0,1\right)$, define $X=\1[\eps_X \leq p_X]$ so that $X \sim B\left(p_X\right)$ is a Bernoulli variable. We consider the special case where $p_X=0.5$. Now, introduce the sigmoid function ${\rm h}\left(x\right) = \left(1+ \exp(-x)\right)^{-1}$, and set, for any $\left(x, v, f\right)\in\{0,1\}^3$ and for a set of real parameters $\alpha_0, \alpha_X, \beta_0, \beta_X, \beta_V, \gamma_0, \gamma_F, \gamma_V $.
\begin{align*}
p_{V}\left(x\right)   &= {\rm h}\left(\alpha_0 + \alpha_X x\right) \\ 
p_{F}\left(x,v\right) &= {\rm h}\left(\beta_0 + \beta_X x + \beta_V v\right) \\ 
p_{A}\left(f,v\right) &= {\rm h}\left(\gamma_0 + \gamma_F f + \gamma_V v\right).
\end{align*}
\noindent Finally, variables $V$, $F$, $A$, $R$ are  defined as
\begin{align*} 
V &=  \1[\eps_V \leq p_{V}\left(X\right)], \\
F &=  \1[\eps_F \leq p_{F}\left(X,V\right)], \\
A &=  \1[\eps_A \leq p_{A}\left(F,V\right)],\\
R &=  F \times A.
\end{align*}

\noindent Moreover, for $\nu \geq 0$, we set
\begin{align*} 
\alpha_0 &=  -\frac{1}{2} \alpha_X \\
\beta_0 &=  -\frac{1}{2}\left(\beta_X+ \beta_V - \nu \right) \\
\gamma_0 &=  -\frac{1}{2}\left(2h\left(\nu\right)\gamma_F + \gamma_V - \nu \right)
\end{align*}
\noindent See Appendix \ref{ap:A04} for details \\

\noindent Then, we compute causal effects and the measure of association on the odds-ratio scale defined below : 
\begin{align*} 
COR\left(X,F|\bW=\bw\right) &= \frac{\p\left(F_1=1 | W=w\right)/\p\left(F_1=0|W=w\right)}{\p\left(F_0=1 | W=w\right)/\p\left(F_0=0| W=w\right)},\\
COR\left(X,R|W=w\right) &= \frac{\p\left(R_1=1 | W=w\right)/\p\left(R_1=0|W=w\right)}{\p\left(R_0=1 | W=w\right)/\p\left(R_0=0| W=w\right)},  \\
OR\left(X,R|W=w,A=1\right) &= OR\left(X,F|W=w,A=1\right) \\
&= \frac{\p\left(F=1 | X=1, W=w, A=1\right)/\p\left(F=0| X=1, W=w, A=1\right)}{\p\left(F=1 | X=0, W=w, A=1\right)/\p\left(F=0| X=0, W=w, A=1\right)}.
\end{align*}
\noindent See Appendix \ref{ap:A05} for the analytic formulas used to compute these effects. \\

\subsection{Numerical results}\label{sec:results}

\begin{figure}[!h]
    \begin{center}
      \includegraphics[scale=0.3]{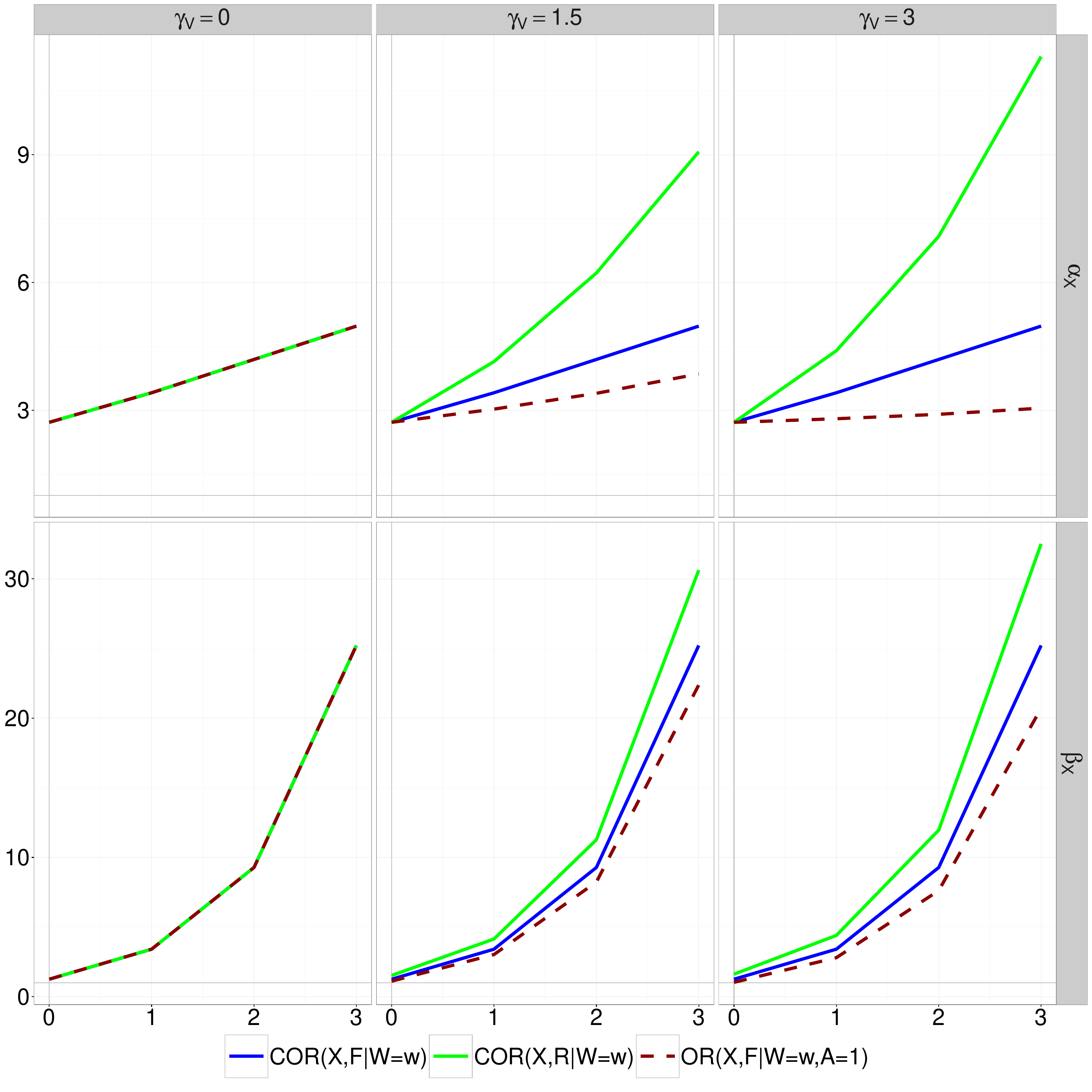} 
    \end{center}
\caption{Causal and associational odds-ratios in the case where $\beta_V = 1$ and $\gamma_F = 4$, and for varying values of the other parameters $\alpha_X, \beta_X, \gamma_V$ . In each panel, along the $x$ axis,  $\alpha_X$ or $\beta_X$ are varied from $0$ to $3$, and the other parameter is set to $1$.} \label{fig:Figure_i}
\end{figure}

\noindent We present results under configurations where $\beta_V = 1$ and $\gamma_F = 4$, because speed $V$ increases the risk of comitting a fault and because $F$ largely increases the risk of having an accident. The choice $\nu = 13$ leads to prevalences of $A$, $R$, and $F$ inferior than $10^{-6}\%$, which can be considered as realistic. Then, we make the three remaining parameters, $\alpha_X$, $\beta_X$ and $\gamma_V$, vary between $0$ and $3$, because alcohol $X$ increases speed $V$ and the risk of committing a fault $F$, and $V$ increases the severity of the accident. Results are presented in Figure \ref{fig:Figure_i}. \\

\noindent In the situation where $\alpha_X \neq 0$ and $\gamma_V \neq 0$, we observe differences between the three quantities $COR\left(X,R|W=w\right)$,  $COR\left(X,F|W=w\right)$ and $OR\left(X,R|W=w,A=1\right)$, and they increase with $\alpha_X$ or $\beta_X$. So, in the general case where $X$ increases the risk to drive fast $V$ and high speed $V$ increases the risk of being involved in a severe crash $A$, $OR\left(X,R|W=w,A=1\right)$ is smaller than the true causal effect $COR\left(X,R|W=w\right)$ in our simulation setting. Finally, the higher the value of $\gamma_V$ and $\alpha_X$ or $\beta_X$, the higher the bias. \\

\noindent When $\gamma_V = 0$, which corresponds to the absence of an arrow between $V$ and $A$ (see case $\left(i\right)$ in Figure \ref{fig7}), we observe no difference between $OR\left(X,R|W=w, A=1\right)$ and $COR\left(X,F|W=w\right)$, and only a tiny one between $COR\left(X,F|W=w\right)$ and $COR\left(X,R|W=w\right)$, whatever the values of $\alpha_X$ or $\beta_X$ (the difference between $COR\left(X,R|W=w\right)$ and $COR\left(X,F|W=w\right)$ is so small can not even be seen on Figure \ref{fig:Figure_i}). This confirms that, in this situation, $COR\left(X,R|W=w\right)\simeq COR\left(X,F|W=w\right)$, so that $COR\left(X,R|W=w\right)$ can be well-approximated by $COR\left(X,F|W=w\right)$. In a setting where $\nu =2$, which is less realistic from the perspective of responsibility analyses since this would correspond to $\p\left(F=1\right)\approx 11\%$ and $\p\left(A=1\right)\approx 11\%$, differences between $COR\left(X,R|W=w\right)$ and $COR\left(X,F|W=w\right)$ get a little larger, while we still have $OR\left(X,R|W=w, A=1\right)=COR\left(X,F|W=w\right)$; see Fig \ref{fig:illustap2} in Appendix \ref{ap:A06}. \\

\noindent Finally, we observe very little differences between $COR\left(X,F|W=w\right)$, $OR\left(X,F|W=w,A=1\right)$ and $OR\left(X,R|W=w,A=1\right)$ for $\alpha_X=0$; again, these differences are so small that they can not be seen from Figure \ref{fig:Figure_i}. In other words, when there is no arrow from $X$ to $V$ but $V$ is a direct cause of $F$ and $A$, despite the fact that the conditional independence $X \indep A | \left(F,\bW\right)$ does not hold, differences between $COR\left(X,F|W=w\right)$, $OR\left(X,F|W=w,A=1\right)$ and $OR\left(X,R|W=w,A=1\right)$ are negligible under the settings that we have considered. These differences become larger for the choice $\nu=2$, and more work would be needed to study how these differences behave under more complex settings involving confounders, interactions, etc. \\

\noindent To recap, under the simple generative model considered here, we can observe that the observable associational effect $OR\left(X,R|W=w,A=1\right)$ approximatively unbiasedly estimates $COR\left(X,R|W=w\right)$ and $COR\left(X,F|W=w\right)$ only when $\gamma_V=0$ or $\alpha_X= 0$. The case where $\gamma_V=0$ is not plausible, because $V$ is always a cause of $A$. The case where $\alpha_X=0$ is plausible, because some causes of $A$ are not causes of $V$. In this latter case, the approximation is only valid under additional but realistic assumptions of low prevalences of $F$ and $A$. Otherwise, $OR\left(X,R|W=w,A=1\right)$ is smaller than the true causal effect under the simple settings considered here where $X$ increases the risk to drive fast $V$. 

\subsection{Practical interest}\label{sec:impli}

\noindent Our results may be useful to discuss estimations derived from responsibility analyses. \\
\noindent In a recent study named ActuSAM, \cite{martin2017cannabis} compare the effect of alcohol consumption and the effect of cannabis intoxication on the risk for being responsible among drivers involved in a fatal crash. The ActuSAM study was made of 2 870 fatal accidents occurring in Metropolitan France in 2011, and of the corresponding 4 059 drivers tested for alcohol and narcotics, and of expert-determined responsibility. A multivariate logistic regression was performed to estimate the effect of alcohol and cannabis on the risk of being responsible for a fatal accident. The association were estimated by odds-ratios and adjusted for age, gender, vehicle category and time of accident. This study concludes that drivers under the influence of alcohol are 17.8 times (12.1-26.1) more likely to be responsible for a fatal accident. Moreover, the higher the blood alcohol concentration, the higher the risk of being responsible for a fatal crash. Regarding cannabis intoxication, the ActuSAM study concludes that drivers under the influence of cannabis multiply their risk of being responsible for causing a fatal accident by 1.65 (1.16-2.34). By comparing the estimations (17.8 vs 1.65), and above all population attributable fractions (PAF) (27.7\% vs 4.2\%), the authors concludes that alcohol consumption remains the main problem on French roads. \\
Our previous results give interesting leads to discuss the estimations derived from ActuSAM study. For instance, in the case of alcohol, it is commonly admitted that alcohol increases the risk to drive fast. We are here in the situation described by Figure \ref{fig7}, where $X$ is a cause of $V$ and $V$ is a cause of $A$. In this situation, our results suggest that the estimation concerning the risk of alcohol would be biased, and probably smaller than the true causal effect. We have completed the ActuSAM results by studying the impact of alcohol, and cannabis, on speed to confirm our reasoning.  Our analysis was based on the same database than the ActuSAM, and on a population of 2 566 drivers. We have included all fatal road crashes which had taken place in France in 2011 in the study, and all the drivers with known speed, alcohol and cannabis consumption. Speed is here considered as continous and we have performed a linear regression model of speed on alcohol and cannabis adjusted on the same set of confounders than the ones choosen in the ActuSAM study. The results are presented in Table \ref{vitesse}. We have knowledge that unbiased estimations are obtained by realising the modelisation of V on the control subpopulation \citep{vanderweele_odds_2010}. Otherwise, collision bias still occurs in the relationship between X and V after conditioning on A. However, there are not enough drunck drivers in the subpopulation of nonresponsible drivers, that is why we use these estimations to comment the ActuSAM results.
 
\begin{table}[h!]
\begin{center}
\caption{Adjusted coefficients for speed driving,(n=2 566, data source Voiesur 2011, fatal accidents)} \label{vitesse} 
\vspace{0.3cm}
    \begin{tabular}{r|c|c}
    Variables  & Coefficients & 95\%CI \\ \hline
    Alcohol $\geq 0.5$ gr/l & 14.12 & [10.44,17.80] \\ \hline
    Cannabis & -3.10 & [-8.03,1.83] \\
    \end{tabular}
\end{center}
\end{table}

\noindent We observe that an alcohol consumption above 0.5 gr/l increases the average speed by 14 km/h; see Table \ref{vitesse}. Even if this estimation may be prone to collider bias as mentioned above, it is line with common knowledge. Since alcohol is likely to increase the risk to drive fast and high speed affects the risk of fatal crash, our previous results suggest that the estimation from the ActuSAM study is biased. More precisely, it would be smaller than the true causal effect of alcohol of being responsible of a fatal crash. \\
Concerning cannabis intoxication, the effect of cannabis on speed has not been established yet. If results presented in Table \ref{vitesse} were valid, it confirms that the relationship between cannabis and speed is not significant, so we are close to DAG illustrating case $\left(iii\right)$. In this situation, our theoretical results suggest that the estimation from ActuSAM study is also biased. Nevertheless, under additional but realistic assumptions of low prevalences of $F$ and $A$, our numerical results suggest that the estimable odds-ratio would be very close to the true causal effect of cannabis.   \\
Our new findings do not negate the global conclusion of the ActuSAM study, because alcohol remains a major health problem on French roads.\\ 

\section{Discussion} \label{sec:ccl}

\noindent In this article, we study responsibility analyses, which are commonly used in the field of road safety epidemiology, under the lense of causal inference. After describing the causal DAG involved in responsibility analyses, we formally show that this type of design does not allow unbiased estimations, even after proper adjustments for confounders. Only one exception exists if pratitioners are interested in the direct effect of $X$ on $R$ (\rm{e.g.} conditioned on $V$) in the situation where  $X$ is not a cause of $V$.
\\

\noindent Focusing first on a binary exposure $X$ which has an impact on $V$, and where $V$ is a cause of a severe accident $A$, we show that $\bw$-specific causal odds-ratios $COR\left(X, R| \bW=\bw\right)$ is not recoverable when data from the most severe accidents only are available. Under additional assumptions, $COR\left(X, R| \bW=\bw\right)$ can be approximated by $COR\left(X,F|\bW=\bw\right)$ if $\left(i\right)$ $V$ was not a cause of $A$, or if $\left(ii\right)$ $X$ was not a cause of $V$ and $V$ a cause of $F$. Nevertheless, none of these cases are plausible in reality. Hence, $COR\left(X, R| \bW=\bw\right)$ is not recoverable and can not be approximated in most situations. We use numerical examples to illustrate our arguments. Under the simple settings that we have considered, we observe that the observable associational effect $OR\left(X,R|W=w,A=1\right)$ is smaller than the true causal effects $COR\left(X, F| \bW=\bw\right)$ or $COR\left(X, R| \bW=\bw\right)$ in the situations where $X$ and $V$, and $V$ and $A$ are positively directly related. We also observe very small differences between the three quantities when $X$ is not a cause of $V$ when prevalences of $F$ and $A$ are low. Our results are useful to interpret recent estimations on the risk of alcohol or cannabis among drivers involved in a fatal crash \citep{martin2017cannabis}. \\

\noindent As a matter of fact, available controls in responsibility analyses are not representative of the non-responsible drivers ($R=0$) when data are restricted to severe accident. Indeed, this population is composed by three types of drivers: the ones for whom $\left(F=0, A=0\right)$, the ones for whom $\left(F=0, A=1\right)$, and the ones for whom $\left(F=1, A=0\right)$. However, because data describes drivers involved in a crash only, the control group is only composed of drivers who did not commit a driving fault but had a severe accident $\left(F=0, A=1\right)$. Rigorously, there is no reason why this subgroup of non-responsible drivers should be similar to the two other subgroups. In particular, drivers involved in a severe crash are likely to drive faster than those who are not involved in an accident, since crash severity is partly caused by speed. Since inappropriate speed is related to alcohol, the subpopulations $\left(F=0, A=0\right)$ and $\left(F=0, A=1\right)$ are  different regarding exposure $X$. Some authors have sensed this non-representativeness issue and have proposed a heuristic transformation of the control group to make the final sample resemble that from a case-control study \citep{LaumonetalBMJ}. However, there is no way to assess whether the final sample is indeed representative without additional information regarding $\p\left(X\right)$ for instance. This type of additional information could also make adjusted causal odds-ratio $COR\left(X,R|\bW=\bw\right)$, causal odds-ratio $COR\left(X,R\right)$ or even $\p\left(R_x=r\right)$, for $\left(x,r\right)\in\{0,1\}$, recoverable from the available data,  even if the structure of the DAG alone does not make these quantities recoverable.  One interesting lead for future research would be to use our formalisation of responsibility analyses to determine which additional information would be sufficient to recover causal effects such as $COR\left(X,R|\bW=\bw\right)$. Note that if we choose an exposure $X$, which has no effect on $V$, $COR\left(X, F| \bW=\bw\right)$ is recoverable and $COR\left(X,R|\bW=\bw\right)$ can be approximated.

\noindent Future research may focus on the formal study of the direction of the selection bias involved responsibility analyses, based on realistic assumptions such as the positivity of the association between $V$ and $A$, $V$ and $F$, etc. Interestingly, when focusing on causal relative risks, rather than causal odds-ratio, we have, for any value $w$ of the confounders,
\begin{align*}
CRR\left(w\right) &= \frac{\p\left(R_1=1| W=w\right)}{\p\left(R_0=1| W=w\right)} \\
&= \frac{\p\left(R=1| X=1, W=w\right)}{\p\left(R=1| X=0, W=w\right)}\\
&= \frac{\p\left(R=1| X=1, W=w, A=1\right)}{\p\left(R=1| X=0, W=w, A=1\right)}\times \frac{\p\left(A=1| X=1\right)}{\p\left(A=1| X=0\right)},
\end{align*}
where we used the fact that $A=0 \Rightarrow R=0.$ Then, the observable relative risk underestimates the causal one if and only if  $\p\left(A=1| X=1\right)/\p\left(A=1| X=0\right)\geq 1$. In particular, this is the case if $X$ and $A$ are positively monotonically associated \citep{vanderweele2010signed}. This could be seen as a reasonable assumption for exposures positively associated with high speed. \\

\noindent In other respect, selection bias is not the only issue in responsibility analyses. First, for drivers involved in a crash, responsibility is usually determined from police reports, from an algorithm \citep{robertson_responsibility_1994, Terhune_1986}, or by experts. This assigned responsibility is a noisy measure of the true responsibility.  In particular, when responsibility is assigned by experts, a safe rule would be to remove all information regarding $X$ from police reports so that experts are blinded to exposure status \citep{brubacher2014culpability}. Otherwise, it is possible that experts use the information regarding $X$ when assigning drivers' responsibility, which is particularly problematic \citep{Salmi_2014}. Second, unmeasured confounders are sources of confounding bias. For instance, risk proneness is likely to be a common cause of alcohol consumption and inappropriate speed, and is therefore a confounder regarding the causal relationship between $X$ and $A$. It is generally unobserved and only loosely related to age, gender, etc, so that adjusting for these covariates is not sufficient and confounding bias arises. \\ 

\noindent Our research enables us to conclude about the validity of responsibility analyses, unfortunately in favor of biased estimations. Nevertheless, this type of design is widely used and no better alternatives exist as of today, so it is important to be able to comment estimations derived from responsibility analyses. In this way, our results give first interesting leads to interpret and discuss these estimations.

\newpage
\bibliographystyle{DeGruyter}
\bibliography{biblio_article1}

\newpage
\begin{appendices}

\section{Causal inference in the Structural Causal Model framework}\label{ap:A0}

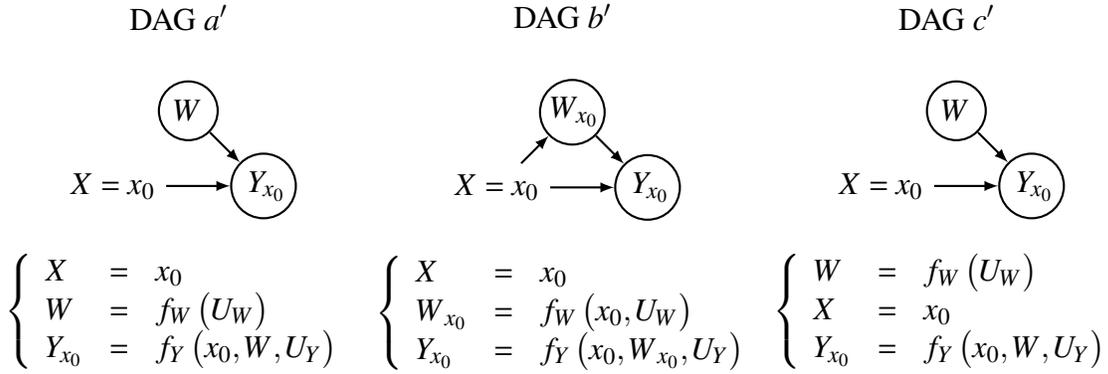
\begin{figure}[!h]
\vspace{0.5cm}
\begin{center} Actual world \end{center}
\begin{minipage}{5cm}
\begin{center} DAG $a$ \end{center}
\begin{center}
\begin{tikzpicture}[scale=1, auto,swap]
\node[var] (X)at(0,0){$X$};
\node[var] (Y)at(2,0){$\,Y$};
\node[var] (W)at(1,1){$\bW$};
\draw[edge] (X)--(Y);
\draw[edge] (W)--(Y);
\end{tikzpicture}
\begin{displaymath}
\left\{ \begin{array}{l c l}
X &=& f_X\left(U_{X}\right) \\
\bW &=& f_{\bW}\left(U_{\bW}\right) \\
Y &=& f_Y\left(X,\bW,U_{Y}\right) 
\end{array} \right.
\end{displaymath}
\end{center}
\end{minipage}
\begin{minipage}{5cm}
\begin{center} DAG $b$ \end{center}
\begin{center}
\begin{tikzpicture}[scale=1, auto,swap]
\node[var] (X)at(0,0){$X$};
\node[var] (Y)at(2,0){$\,Y$};
\node[var] (W)at(1,1){$\bW$};
\draw[edge] (X)--(Y);
\draw[edge] (X)--(W);
\draw[edge] (W)--(Y);
\end{tikzpicture}
\begin{displaymath}
\left\{ \begin{array}{l c l}
X &=& f_X\left(U_{X}\right) \\
\bW &=& f_{\bW}\left(X, U_{\bW}\right) \\
Y &=& f_Y\left(X,\bW,U_{Y}\right) 
\end{array} \right.
\end{displaymath}
\end{center}
\end{minipage}
\begin{minipage}{5cm}
\begin{center} DAG $c$ \end{center}
\begin{center}
\begin{tikzpicture}[scale=1, auto,swap]
\node[var] (X)at(0,0){$X$};
\node[var] (Y)at(2,0){$\,Y$};
\node[var] (W)at(1,1){$\bW$};
\draw[edge] (X)--(Y);
\draw[edge] (W)--(X);
\draw[edge] (W)--(Y);
\end{tikzpicture}
\begin{displaymath}
\left\{ \begin{array}{l c l}
\bW &=& f_{\bW}\left(U_{\bW}\right) \\
X &=& f_X\left(\bW,U_{X}\right) \\
Y &=& f_Y\left(X,\bW,U_{Y}\right) 
\end{array} \right.
\end{displaymath}
\end{center}
\end{minipage}

\vspace{0.3cm}
\begin{center} Corresponding counterfactual worlds, following physical intervention $do\left(X=x_0\right)$ \end{center}
\begin{minipage}{5cm}
\begin{center} DAG $a'$ \end{center}
\begin{center}
\begin{tikzpicture}[scale=1, auto,swap]
\node[varf] (X)at(0,0){$X=x_0$};
\node[var] (Y)at(2,0){$Y_{x_0}$};
\node[var] (W)at(1,1){$\bW$};
\draw[edge] (X)--(Y);
\draw[edge] (W)--(Y);
\end{tikzpicture}
\begin{displaymath}
\left\{ \begin{array}{l c l}
X &=& x_0 \\
\bW &=& f_{\bW}\left(U_{\bW}\right) \\
Y_{x_0} &=& f_Y\left(x_0,\bW,U_{Y}\right) 
\end{array} \right.
\end{displaymath}
\end{center}
\end{minipage}
\begin{minipage}{5cm}
\begin{center} DAG $b'$ \end{center}
\begin{center}
\begin{tikzpicture}[scale=1, auto,swap]
\node[varf] (X)at(0,0){$X=x_0$};
\node[var] (Y)at(2,0){$Y_{x_0}$};
\node[var] (W)at(1,1){$\!\!\bW_{\!x_0}$};
\draw[edge] (X)--(Y);
\draw[edge] (X)--(W);
\draw[edge] (W)--(Y);
\end{tikzpicture}
\begin{displaymath}
\left\{ \begin{array}{l c l}
X &=& x_0 \\
\bW_{x_0} &=& f_{\bW}\left(x_0, U_{\bW}\right) \\
Y_{x_0} &=& f_Y\left(x_0,\bW_{x_0},U_{Y}\right) 
\end{array} \right.
\end{displaymath}
\end{center}
\end{minipage}
\begin{minipage}{5cm}
\begin{center} DAG $c'$ \end{center}
\begin{center}
\begin{tikzpicture}[scale=1, auto,swap]
\node[varf] (X)at(0,0){$X=x_0$};
\node[var] (Y)at(2,0){$Y_{x_0}$};
\node[var] (W)at(1,1){$\bW$};
\draw[edge] (X)--(Y);
\draw[edge] (W)--(Y);
\end{tikzpicture}
\begin{displaymath}
\left\{ \begin{array}{lc l}
\bW &=& f_{\bW}\left(U_{\bW}\right) \\
X &=& x_0 \\
Y_{x_0} &=& f_Y\left(x_0,\bW,U_{Y}\right) 
\end{array} \right.
\end{displaymath}
\end{center}
\end{minipage}
\caption{Examples of DAGs and SEMs: actual and counterfactual worlds.}
\label{fig3}
\end{figure}

This Appendix presents the basis of causal inference, with a special emphasis on the SCM framework \citep{Pearl_2000}. We consider the same simple setting as that of Section 2 where cause $X$ and effect $Y$ are binary, and vector $\bW$ is a set of categorical variables. A first useful tool to describe causal assumptions underlying a causal model $\cM$ is the DAG. See the three examples considered in Figure \ref{fig3}. Under each of these examples, the DAG can be translated into a Structural Equation Model (SEM), that is a set of three structural equations, involving three autonomous functions $f_X, f_{\bW}$ and $f_Y$, as well as three exogenous random variables $U_X, U_{\bW}$ and $U_Y$, sometimes called disturbances \citep{Pearl_2000,Pearl_2009}. 
We will assume that $U_X, U_{\bW}$ and $U_Y$ are jointly independent. These sets of structural equations govern the distributions of the variables $\bcV = \left(X, \bW, Y\right)$ observed in the actual world and available in observational studies. But they also allow the description of what we would have observed in counterfactual worlds, if we could have intervened and forced the exposure to be set to $X=x_0$ (with $x_0\in\{0,1\}$ in the simple scenario considered here). This physical intervention can be modeled by the \textit{do}-operator introduced by \cite{Pearl_1995,Pearl_2000}. In model $\cM$, this operator modifies the value of $X$: instead of being the result of the autonomous function $f_X$, it is set to $X=x_0$. The $do$-operator further removes any arrow pointing to $X$ in the DAG. For instance, in Figure \ref{fig3}, DAG $a'$ and DAG $c'$ are identical while DAG $a$ and DAG $c$ are different: the two causal models are identical in the counterfactual world that we would observe after the intervention $do\left(X=x_0\right)$, but are different in the actual world. 

\textit{Causal effects} represent  a general ability to transfer changes among covariates \citep{Pearl_2009}, and the $do$-operator is a key tool for their definition in the SCMs framework. The intervention $do\left(X=x_0\right)$ does not affect functions $f_{V_1}$ (nor disturbances $U_{V_1}$) for $V_1 \in \bV\setminus \{X\}$, but it does affect the distribution of descendants of $X$ in the DAG (its children, the children of its children, etc.). More precisely, in the {\em counterfactual} world following the intervention $do\left(X=x_0\right)$, model $\cM$ would be model $\cM_{x_0}$ leading to observations of the variables $\bcV\left(x_0\right) = \left(x_0, \bW_{x_0}, Y_{x_0}\right)$ instead of $\bcV = \left(X, \bW, Y\right)$. In case (a) for instance, we have $\bW_{x_0} = f_{\bW}(U_{\bW}) = \bW$ but $Y_{x_0} = f_Y\left(x_0, \bW_{x_0}, U_Y\right) = f_Y\left(x_0, \bW, U_Y\right)$, which is typically different from $Y$, unless $X=x_0$. The random variables $\left(Y_{x_0}\right)_{x_0\in\{0,1\}}$ are not (fully) observed: they are counterfactual variables, and correspond to potential outcomes \citep{splawa-neyman_application_1990,rubin_estimating_1974,holland_statistics_1986}. In the Marginal Structural Models (MSMs) framework \citep{Robins_2000},  these counterfactual variables $\left(Y_{x_0}\right)_{x_0\in\{0,1\}}$ are connected to the variable $Y$ observed in the actual world, through consistency constraints like the coherence assumption \citep{Robins_1986}: 
\begin{equation*}
\text{(Coh)} \quad\quad \left(X=x\right) \Rightarrow \left(Y = Y_x\right), \quad \text{for\ all\ potential\ values\ } x\ \text{of}\ X.
\end{equation*}
In words, this assumption states that the actual outcome for an individual whose actual level of exposure is $x$ equals the outcome we would observe for this same individual in the counterfactual world following $do\left(X=x\right)$, {\em i.e.}, where exposure would be ``physically'' set to value $x$. This assumption can be violated in randomized clinical trial if full adherence is not guaranteed for instance. In the simple case of binary exposure considered here, the coherence assumption entails that $Y = X Y_1 + \left(1-X\right) Y_0$. Under this coherence assumption, $Y_1$ and $Y_0$ are therefore partially observed and causal inference from observational data can be seen as a missing data problem. One first advantage of SCMs over MSMs rests in the precise definition of the counterfactual variables from the model, thanks to the DAG and the corresponding SEM. For instance, it is easy to show that the coherence assumption is automatically fulfilled in the structural interpretation of counterfactual variables considered here, since $Y = f_Y\left(X, \bW, U_Y\right)$ and $Y_x = f_Y\left(x, \bW,U_Y\right)$. 

Once potential outcomes have been introduced (and precisely defined in the SCM framework), causal effects can be  precisely defined. In particular, in the simple example considered here where both $X$ and $Y$ are binary, the average causal effect of $X$ on $Y$ is defined as 
\begin{equation}
{\rm ACE}= \E\left(Y_{1} - Y_{0}\right) = \p\left(Y_1 = 1\right) - \p\left(Y_0 = 1\right). \label{eq:ACE}
\end{equation}
In the SCM framework, $ \p\left(Y_x = 1\right)$ is the probability that the outcome variable $Y$ would equal 1 in the counterfactual world following the intervention $do\left(X=x\right)$: using Pearl's notations, we have $\p\left(Y_x = 1\right) = \p\left(Y=1| do\left(X=x\right)\right)$ (see, {\em e.g.}, Equation (7) in \cite{Pearl_2009}). The average causal effect measures the difference between the risk of observing the outcome in the counterfactual world where all individuals would be exposed, $\p\left(Y_1 = 1\right)=\p\left(Y=1 | do\left(X=1\right)\right)$, and the risk of observing the outcome in the counterfactual world where exposure would be eliminated, $\p\left(Y_0 = 1\right)=\p\left(Y=1 | do\left(X=0\right)\right)$. Here, causal effect is defined on the excess risk scale, but causal risk ratio or odds-ratio can of course be defined similarly: e.g., for the causal risk ratio, it is defined as $\p\left(Y_1 = 1\right) / \p\left(Y_0 = 1\right)$. 

So far, we have shown that the introduction of the $do$-operator and the counterfactual variables $Y_1$ and $Y_0$ allow a precise definition of causal effects. However, because such causal effects rely on quantities that are not (fully) observed in the actual world, a natural question arises whether these effects can be estimated from observational data.

Consider a general DAG $G$ composed of observed variables ${\boldsymbol {\cal V}}$, with $\left(X, Y\right)\in {\boldsymbol {\cal V}}$. In the SCMs framework, $\p\left(Y=y|do\left(X=x\right)\right) = \p\left(Y_x = y\right)$ is said to be identifiable if the assumptions embedded in DAG $G$ ensure that this quantity is expressible in terms of the observable distribution  $\p\left({\boldsymbol {\cal V}}=\bv\right)$; see Definition 1 in \cite{Bareinboim_Tian_2015} for instance. 

Here we start by recalling standard conditions introduced in the literature, that are sufficient for identifiability of $\p\left(Y=y|do\left(X=x\right)\right) = \p\left(Y_x = y\right)$, when combined with the coherence assumption (Coh). These conditions ensure a certain level of ``comparability'' between exposed and non-exposed individuals in the actual world. In particular, the ignorability assumption reads as follows \citep{greenland_causal_1999}:
\begin{equation*}
\text{(Ign)} \quad\quad Y_x \indep X, \quad \text{for\ all\ potential\ values\ } x\ \text{of}\ X,
\end{equation*}
where $V_1\indep V_2$ stands for ``$V_1$ and $V_2$ are independent''. This assumption states that the outcome $Y_x$ we would observe in the counterfactual world where exposure would be set to value $x$ is independent of the actual level of exposure $X$. Under (Ign) and (Coh), the causal effect of $X$ can be estimated from observations of the actual world, as long as $0<\p\left(X=1\right)<1$. Indeed, by successively applying assumption (Ign) and (Coh), it follows that 
\begin{align}
\p\left(Y_x = 1\right) = \p\left(Y_x = 1 | X=x\right) = \p\left(Y = 1 | X=x\right) \label{eq:Causal_Ign}
\end{align}
Under (Coh) and (Ign) the association measure between $X$ and $Y$  is said to be unconfounded and equals the causation measure. For instance, under (Coh) and (Ign),  $\p\left(Y = 1 | X=1\right)-\p\left(Y = 1 | X=0\right) = \p\left(Y_1 = 1\right)  - \p\left(Y_0 = 1\right)$. Of course, (Ign) is a strong assumption and, as will be made clearer below, it does not hold in the presence of confounders for instance. In such cases, a conditional version of (Ign) can also been considered \citep{Rosenbaum_Rubin_1983}. In particular, assume that, for some variable $\bW$, the following assumption holds:
\begin{equation*}
\text{(C.Ign)} \quad\quad Y_x \indep X  | \bW , \quad \text{for\ all\ potential\ values\ } x\ \text{of}\ X,
\end{equation*}
where $V_1\indep V_2 | V_3$ stands for ``$V_1$ and $V_2$ are conditionally independent given $V_3$''. Then the so-called adjustment formula, or back-door formula \citep{Pearl_1995}, holds too: 
\begin{align*}
\p\left(Y_x = 1\right) &= \sum_{\bw} \p\left(Y_x = 1 | \bW=\bw\right) \p\left(\bW=\bw\right)\\
& =  \sum_{\bw} \p\left(Y_x = 1 |X=x, \bW=\bw\right) \p\left(\bW=\bw\right) \\
&=  \sum_{\bw} \p\left(Y = 1 | X=x, \bW=\bw\right) \p\left(\bW=\bw\right). 
\end{align*}
Therefore, under (Coh) and (C.Ign), we have
\begin{equation}
\text{ACE} = \sum_{\bw}  \{\p\left(Y = 1 | X=1, \bW=\bw\right) - \p\left(Y = 1 | X=0, \bW=\bw\right)\}\p\left(\bW=\bw\right),\label{eq:ACE_Cond}
\end{equation}
and ACE corresponds to a marginalized version of the adjusted excess risk. Under the Experimental Treatment Assignment (ETA) assumption, that is  $0<\p\left(X=x|\bW\right)<1$ almost everywhere, ACE can be estimated from observations of the actual world. Methods adapted to situations where the ETA assumption fails to hold have also been proposed (see \cite{Moore_2012} for a review and further extensions). 

\newpage
\section{Proof of approximation (\ref{eq:CORCRR})}\label{ap:CORCRR}

Under the assumption of no unmeasured confounder, $OR\left(X, F| \bW=\bw\right) = COR\left(X, F| \bW=\bw\right)$. Moreover, if  $\p\left(F_x=1| \bW=\bw\right)$ is small for $x\in\{0,1\}$, then $COR\left(X, F| \bW=\bw\right)\approx CRR\left(X, F| \bW=\bw\right)$, while if $\p\left(R_x=1|\bW=\bw\right)$ is small for $x\in\{0,1\}$, then $COR\left(X, R| \bW=\bw\right)\approx CRR\left(X, R| \bW=\bw\right)$. Therefore, under these assumptions, and if $X\indep A|\left(F, \bW\right)$, we have
\begin{align*}
COR\left(X, R| \bW=\bw\right)&\approx CRR\left(X, R| \bW=\bw\right)\\
& = \frac{\p\left(R_1 =1| \bW=\bw\right)}{\p\left(R_0 =1| \bW=\bw\right)}\\
&= \frac{\p\left(F_1 =1, A_1=1| \bW=\bw\right)}{\p\left(F_0 =1, A_0=1| \bW=\bw\right)}\\
&= \frac{\p\left(F_1 =1, A_1=1| X=1, \bW=\bw\right)}{\p\left(F_0 =1, A_0=1| X=0, \bW=\bw\right)}\\
&= \frac{\p\left(F =1, A=1| X=1, \bW=\bw\right)}{\p\left(F =1, A=1| X=0, \bW=\bw\right)}\\
&= \frac{\p\left( A=1| F =1,X=1, \bW=\bw\right)\p\left(F =1|X=1, \bW=\bw\right)}{\p\left(A=1| F =1, X=0, \bW=\bw\right)\p\left(F =1|X=0, \bW=\bw\right)}\\
&= \frac{\p\left( A=1| F =1, \bW=\bw\right)\p\left(F =1|X=1, \bW=\bw\right)}{\p\left(A=1| F =1,\bW=\bw\right)\p\left(F =1|X=0, \bW=\bw\right)}\\
&= \frac{\p\left(F =1|X=1, \bW=\bw\right)}{\p\left(F =1|X=0, \bW=\bw\right)}\\
&= \frac{\p\left(F_1 =1|\bW=\bw\right)}{\p\left(F_0 =1|\bW=\bw\right)}\\
&= CRR\left(X, F| \bW=\bw\right)\\
&\approx COR\left(X, F| \bW=\bw\right)
\end{align*}

\section{Formal definition of $OR\left(X,R|W=w,A=1\right)$} \label{ap:A03}

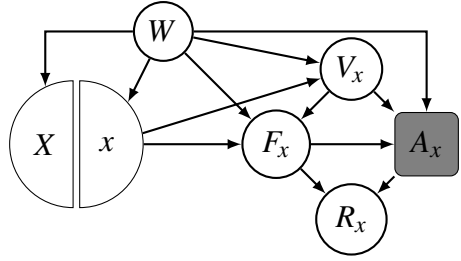
\begin{figure}[!h]
\vspace{0.3cm}
\begin{center}
\begin{tikzpicture}[scale=1, auto,swap]
\node[varleft] (X)at(-0.07,0){$X$};
\node[varright] (x)at(0.75,0){$x$};
\node[varr] (F)at(3,0){$F_x$};
\node[varCond] (A)at(5,0){$A_x$};
\node[var] (W) at (1.5,1.5) {$\bW$};
\node[var] (V) at (4,1) {$V_x$};
\node[varr] (R) at (4,-1) {$R_x$};
\draw[thick,->,>=latex] (W) -| (X);
\draw[edge] (W)--(x);
\draw[edge] (W)--(V);
\draw[edge] (W)--(F);
\draw[thick,->,>=latex] (W) -| (A);
\draw[edge] (x)--(F);
\draw[edge] (F)--(A);
\draw[edge] (F)--(R);
\draw[edge] (A)--(R);
\draw[edge] (x)--(V);
\draw[edge] (V)--(F);
\draw[edge] (V)--(A);
\end{tikzpicture}
\end{center}
\caption{The SWIT resulting from the intervention $do\left(X=x\right)$ in case \textit{(ii)}}
\label{fig5}
\end{figure}

Denote by $R_x$, $F_x$, $V_x$, $A_x$ the counterfactual outcome variables we would observe in the counterfactual world following the intervention $X=x$. Figure \ref{fig5} presents the SWIT corresponding to the intervention $do\left(X=x\right)$ \citep{SWIGs}.

\noindent From this representation, it directly follows that $R_x \indep X | W$, and therefore that $R_x \indep X | \left(\bW,A_x\right)$. Then, the following holds 
\begin{align}
 \p\left(R = 1 | X=x, \bW=\bw, A=1\right) &= \p\left(R_x = 1 | X=x, \bW = \bw, A_x=1\right) \quad {\rm by\ consistency}\nonumber \\
 &= \p\left(R_x = 1 | \bW = \bw, A_x=1\right)\quad {\rm since\ } R_x \indep X | \left(\bW,A_x\right).\label{eq:der}
\end{align}

Consequently, it can be shown that the adjusted odds-ratio conditioned on $A=1$ is
\begin{align*}
OR\left(X, R | \bW=\bw, A=1\right) & = \frac{\p\left(R_1=1 | \bW=\bw, A_1=1\right)/\p\left(R_1=0 | \bW=\bw, A_1=1\right)}{\p\left(R_0=1 | \bW=\bw, A_0=1\right)/\p\left(R_0=0 | \bW=\bw, A_0=1\right)}.
\end{align*}

\section{Choice of constant in numerical illustration} \label{ap:A04}

We note $\p\left(X=1\right)=p_X, \p\left(V=1\right)=p_V$ and $\p\left(F=1\right)=p_F$. Inspired by \cite{sperrin_collider_2016}, we set
\begin{align*} 
\alpha_0 &=  - \left(p_X \alpha_X\right) \\
\beta_0 &=  - \left(p_X \beta_X + p_V \beta_V - \nu' \right) \\
\gamma_0 &=  -\left(p_F \gamma_F + p_V \gamma_V - \nu' \right)
\end{align*}
\noindent with $p_X = p_V = 0.5$ and $p_F=h\left(\nu\right)$, so that the prevalence of $V$ remains close to 50\% and the prevalences of $F$ and $A$ remain close to $h\left(\nu\right)$. Consequently,
\begin{align*} 
\alpha_0 &=  -\frac{1}{2} \alpha_X \\
\beta_0 &=  -\frac{1}{2}\left(\beta_X+ \beta_V - \nu \right) \\
\gamma_0 &=  -\frac{1}{2}\left(2h\left(\nu\right) \gamma_F + \gamma_V - \nu \right)
\end{align*}

\section{Analytic expression used to compute causal and associational effect in numerical illustration}\label{ap:A05}

\noindent Under our generative model, we get, for $x \in \lbrace 0,1 \rbrace $,
\begin{align*} 
\p\left(F_x=1 | W=w\right) &= \p\left(F=1 | X=x, W=w\right)\\
&= \sum_{v \in \left(0,1\right)}\lbrace \p\left(F=1 | X=x, V=v, W=w\right) \p\left(V=v | X=x, W=w\right)\rbrace \\
&= {\rm h}\left(\beta_0 + \beta_X x + \beta_V\right) {\rm h}\left(\alpha_0 + \alpha_X x\right) + {\rm h}\left(\beta_0 + \beta_X x\right)  {\rm h}\left(-\left(\alpha_0 + \alpha_X x\right)\right)
\end{align*}

\begin{align*} 
\p\left(R_x=1 | W=w\right) &= \p\left(A=1,F=1 | W=w\right)\\
&= \sum_{v \in \left(0,1\right)} \lbrace \p\left(A=1 | F=1, V=v, W=w\right) \p\left(F=1 | X=x, V=v, W=w\right) \\
&\quad\quad\quad\quad\quad  \p\left(V=v | X=x, W=w\right)\rbrace \\
&= {\rm h} \left(\gamma_0 + \gamma_F + \gamma_V\right) {\rm h} \left(\beta_0 + \beta_X x + \beta_V\right) {\rm h} \left(\alpha_0 + \alpha_X x\right) + \\ &\quad\quad  {\rm h} \left(\gamma_0 + \gamma_F\right) {\rm h} \left(\beta_0 + \beta_X x\right)  {\rm h} \left(-\left(\alpha_0 + \alpha_X x\right)\right)
\end{align*}

\begin{align*} 
&\p\left(F=1 | X=x, W=w, A=1\right) \\
&= \frac{\p\left(F=1, X=x, W=w, A=1\right)}{\p\left(X=x, W=w, A=1\right)} \\
&= \frac{\p\left(A=1| F=1, X=x, W=w\right)\p\left(F=1 X=x, W=w\right) }{\p\left(A=1| X=x, W=w\right)\p\left(X=x| W=w\right)\p\left(W=w\right)} \\
&= \frac{\sum\limits_{v \in \left(0,1\right)} \lbrace \p\left(A=1| F=1, V=v, W=w\right) \p\left(F=1| X=x, V=v, W=w\right) \p\left(V=v| X=x, W=w\right) \rbrace}{\sum\limits_{v,f \in (\left(0,1\right)}^{\textcolor{white}{n}} \lbrace \p\left(A=1| F=f, V=v, W=w\right) \p\left(F=1| X=x, V=v, W=w\right) \p\left(V=v| X=x, W=w\right) \rbrace} \\
&= {\rm h} \left(\gamma_0 + \gamma_F + \gamma_V\right) {\rm h} \left(\beta_0 + \beta_X x + \beta_V\right) {\rm h} \left(\alpha_0 + \alpha_X x\right) + {\rm h} \left(\gamma_0 + \gamma_F\right) {\rm h} \left(\beta_0 + \beta_X x\right)  {\rm h} \left(-\left(\alpha_0 + \alpha_X x\right)\right) \times \\
& \quad\quad [{\rm h} \left(\gamma_0 + \gamma_F + \gamma_V\right) {\rm h} \left(\beta_0 + \beta_X x + \beta_V\right) {\rm h} \left(\alpha_0 + \alpha_X x\right) + {\rm h} \left(\gamma_0 + \gamma_F\right) {\rm h} \left(\beta_0 + \beta_X x\right)  {\rm h} \left(-\left(\alpha_0 + \alpha_X x\right)\right)+ \\
&\quad\quad {\rm h} \left(\gamma_0  + \gamma_V \right) {\rm h} \left(-\left(\beta_0 + \beta_X x + \beta_V\right)\right) {\rm h} \left(\alpha_0 + \alpha_X x\right) +  {\rm h} \left(\gamma_0\right)  {\rm h} \left(-\left(\beta_0 + \beta_X x\right)\right) {\rm h} \left(-\left(\alpha_0 + \alpha_X x\right)\right)]^{-1}
\end{align*}

\section{Numerical illustration with higher prevalences of $F$ and $A$}\label{ap:A06}

\begin{figure}[!h]
    \begin{center}
      \includegraphics[scale=0.3]{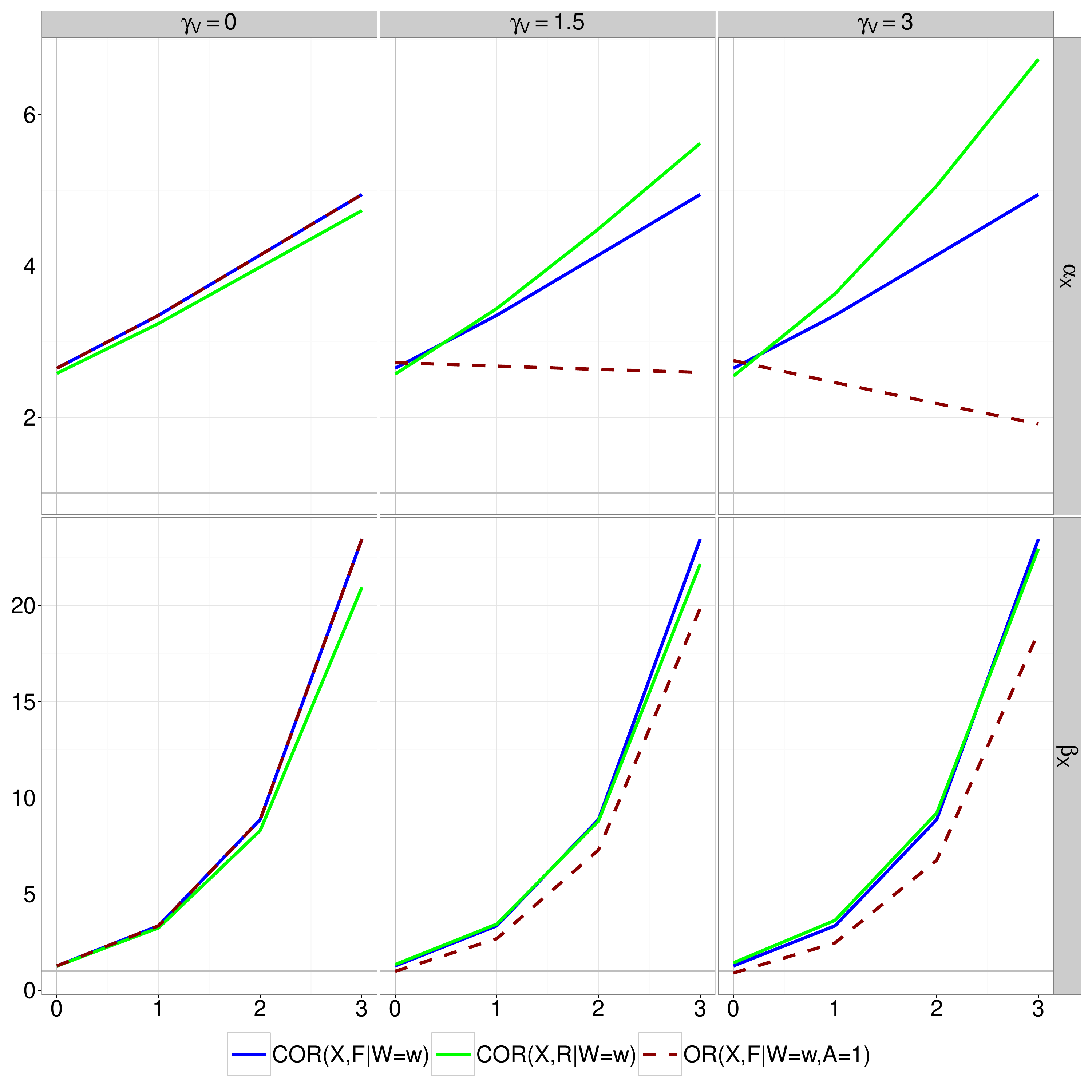} 
    \end{center}
\caption{Causal and associational odds-ratios in the case where $ \nu=2$, and $\beta_V = 1$ and $\gamma_F = 4$, and for varying values of the other parameters $\alpha_X, \beta_X, \gamma_V$ . In each panel, along the $x$ axis,  $\alpha_X$ or $\beta_X$ are varied from $0$ to $3$, and the other parameter is set to $1$.} \label{fig:illustap2}
\end{figure}

\noindent Here, we consider the same causal model and the same set of values for the different paramaters considered in the model. We only change the value of $\nu$ from $13$ to $2$ to get higher prevalences of $F$ and $A$. \\
With $\nu=2$, the prevalence of $F$ is around $14\%$ and the prevalence of $A$ is around $19\%$. It is not realistic in our case but it allows us to illustrate the presence of bias when $\alpha_X = 0$ (See Figure \ref{fig:illustap2}).

\end{appendices}
\end{document}